
\magnification=1200
\parskip=10pt
\parindent=14pt
\baselineskip=18pt
\input mssymb
\pageno=0
\footline={\ifnum \pageno <1 \else \hss \folio \hss \fi }
\line{\hfil{DAMTP-R/94/15}}
\line{\hfil{March, 1994}}
\vskip .4in
\centerline{\bf DIVERGENCES IN THE MODULI SPACE
INTEGRAL AND}
\vskip 1pt
\centerline{\bf ACCUMULATING HANDLES IN THE INFINITE-GENUS LIMIT}
\vskip .6in
\centerline {Simon Davis}
\vskip .5in
\centerline {International Centre for Theoretical Physics}
\vskip 1pt
\centerline {Strada Costiera 11, 34100 Trieste, Italy }
\vskip 5pt
\centerline{and}
\vskip 5pt
\centerline{Department of Applied Mathematics and Theoretical Physics}
\vskip 1pt
\centerline{University of Cambridge}
\vskip 1pt
\centerline{Silver Street, Cambridge CB3 9EW {\footnote{*}{Present address}}}
\vskip .55in
{\bf Abstract}.  The symmetries associated with the closed bosonic string
partition function are examined so that the integration region in Teichmuller
space can be determined.  The conditions on the period matrix defining the
fundamental region can be translated to relations on the parameters of the
uniformizing Schottky group.  The growth of the lower bound for the regularized
partition function is derived through integration over a subset of the
fundamental region.
\vfill
\eject

\noindent{\bf 1. Introduction}

The elimination of infrared divergences in scattering amplitudes of superstring
theories promises a consistent quantum theory including gravity as part of the
low-energy limit.  An understanding of space-time at the most fundamental level
could be achieved with the development of such a theory.

This requires a complete formulation of the theory based on the sum over
string histories arising in the path integral.  Both the entire perturbative
amplitude, and possibly non-perturbative effects, could be obtained in this
approach.

Interest has centered recently on a divergence in bosonic string theory that
arises from summing the contributions at each loop, which have been rendered
finite individually through a genus-independent regularization [1].  The
cut-off
introduced for the bosonic string excludes effectively closed surfaces of
sufficiently large genus, but it will be shown that surfaces having an ideal
boundary with positive linear measure do arise in the sum over all orders in
the perturbation expansion.  The source of the divergence can then be traced to
these surfaces, which may be interpreted as representing a non-perturbative
effect in string theory.

The counting of these surfaces, and the effectively closed surfaces at higher
genus, in the path integral could impact on the finiteness properties of the
superstring path integral.

The investigation of the divergence begins here with a study of the measure and
the domain in the integral for the partition function.  The measure is derived
from the path integral weighting factor and a choice of coordinates on the
the moduli space of metrics.  The two most frequently used measures are those
defined by the light-cone and Polyakov approaches, which require manifest
unitarity or covariance of the string theory respectively.  The light-cone
diagram is constructed so that the momentum of the external string is
proportional to the distance between the cuts at initial and final times,
while the internal cuts correspond to the joining and splitting of strings.
The conformal mapping from the string diagram to a planar domain with disks
removed and punctures at the positions of the vertex operators transforms
paths from the boundary of the diagram to the internal cuts and the paths
around the cuts to a-cycles and b-cycles respectively [2].  The planar domains
with 2g disks removed are the Schottky covering surfaces for a Riemann surface
of genus g, and by the retrosection theorem, every compact Riemann surface
can be uniformized by a Schottky group generated by g Mobius transformations
[3].

The path integral in the Euclidean formulation is
$$\sum_{g=0}^{\infty}~\kappa^g~\int~{{D[h_{\alpha\beta}]~D[X^{\mu}]}\over
{Vol(Diff~ \Sigma_g)~Vol(Conf~\Sigma_g)}}~e^{-{1\over 2}\int_{\Sigma_g}
d^2 \xi \sqrt{h} h^{\alpha\beta}
\partial_{\alpha}X^{\mu}\partial_{\beta}X_{\mu}}
\eqno(1)$$
The integration over the space of metrics is reduced first to the 3g-3
complex-dimensional Teichmuller space $T_g$ by factoring by the identity
component of the diffeomorphism group $Diff_0(\Sigma_g)$, and then it can be
reduced further to the moduli space $M_g$ by factoring by the mapping class
group $Diff(\Sigma_g)/Diff_0(\Sigma_g)$.  Since a compact Riemann surface
of genus g can be represented as $D/\Gamma$, where $\Gamma$ is a Schottky group
with g generators and D is the set of ordinary points of $\Gamma$ [the
complement of the set of limit points in the extended complex plane], the
space of inequivalent complex structures on the surface can be coordinatized
by the multipliers $K_n, n=1,...,g$ and fixed points $\xi_{1n},\xi_{2n},
n=1,...,g$ of the generating Mobius transformations $T_1,...,T_g$, with three
of the parameters fixed by an overall PSL(2,${\Bbb C}$) transformation.

The measure on moduli space can then be expressed [4] as
$$\eqalign{\prod_{m=1}^g~{{d^2 K_m}\over {\vert K_m \vert^4} }\vert 1 -
K_m\vert^4
{1\over {Vol (SL(2,{\Bbb C}))}} &\prod_{n=1}^g~{{d^2 \xi_{1n} ~ d^2\xi_{2n}}
\over {\vert \xi_{1n}-\xi_{2n}\vert^4}} [det(Im~\tau)]^{-13g}
\hfill
\cr
\prod_{\alpha}~^{\prime}&\prod_{p=1}^{\infty} \vert 1-K_{\alpha}^p\vert^{-48}
\prod_{\alpha}~^{\prime} \vert 1 - K_{\alpha}\vert^{-4}\cr}
\eqno(2)$$
where $\prod_{\alpha}^{\prime}$ is a product over conjugacy classes of
primitive  elements of the group $\Gamma$ and $\tau$ is the period matrix [5]
$$\tau_{mn}~=~{1\over {2 \pi i}}\left[ ln K_m \delta_{mn}~+~
\sum_{\alpha}~^{(m,n)}~ln \left({{\xi_{1m}-V_{\alpha} \xi_{1n}}\over {\xi_{1m}
-V_{\alpha} \xi_{2n}}}{{\xi_{2m}-V_{\alpha}\xi_{2n}}\over {\xi_{2m}-V_{\alpha}
\xi_{1n}}}\right)~\right]
\eqno(3)$$
with $\sum_{\alpha}^{(m,n)}$ excluding elements with $T_m^{\pm 1}$ as the
left-most member and $T_n^{\pm 1}$ as the right-most member.

An alternative expression for the partition function can be obtained
through the evaluation of the determinant factors giving rise to Selberg
trace functions.  The equivalence of the two approaches has been demonstrated
by relating the determinants of these operators and path integrals over
the positions of interaction points [6].  As the equivalence should extend to
measures, the differential volume element in equation (2) can be used to study
the behaviour of the partition function with respect to the genus.  Moreover,
the Schottky group formalism has been found to be one of the few methods useful
for calculating physically relevant higher-genus scattering amplitudes [7][8].

It has been established that the string integrand is invariant with respect
to the symplectic modular group [9], which is associated with a change of
homology basis.  When the a-cycles and b-cycles transform as
$$\left({{a^{\prime}}\atop {b^{\prime}}}\right)~=~
\left(\matrix{A&B\cr
              C&D\cr}\right) \left({a\atop b}\right)~~~~~~~
\left(\matrix{A&B\cr
              C&D\cr}\right) ~\in~ Sp(2g;{\Bbb Z})
\eqno(4)$$
the canonical basis of one-forms transforms as
$$[\tilde\omega_1,...,\tilde\omega_g]~=~[\omega_1,...,\omega_g](A+B\tau)^{-1}
\eqno(5)$$
and the period matrix transforms as
$$\tau ~\to~ (C+D \tau)(A+B \tau)^{-1}
\eqno(6)$$
The transformations are induced by diffeomorphisms not connected to the
identity.  A particular set of mappings, Dehn twists, are defined by cutting
out cylinders in the surface, twisting the cylinder around a chosen cycle,
and gluing the cylinder back to the surface [10].  These twists mix the
a-cycles
and b-cycles as given in equation (4) and may generate transformations that
are not connected to the identity compoment of the diffeomorphism group.
As the integer entries of the symplectic matrix M are restricted by the
condition that the determinant should equal 1 and that the intersection matrix
 is preserved $M^T JM=J$, the image of the symplectic group in the
mapping class group must be supplemented by the transformations not acting
on the homology basis, the Torelli group [11].

Integration over the Schottky group parameters, or equivalently the period
matrix elements, through equation (3), already implies factorization by
the Torelli group.  An integration region in the space of period matrices can
be defined using the action of the symplectic modular group in equation (6).
Transformations in the Schottky group space $\{ K_n, \xi_{1n}, \xi_{2n}\}$
also lead to a mixing of a- and b- cycles, because configurations in which
the isometric circles of the fundamental generators are tangent allow for
paths which are combinations of the original a- and b-cycles.  Thus, an
integral
over a region in parameter space is only equivalent to a lower bound for
the moduli space integral when the region lies inside the fundamental domain
of the modular group.  The estimate of the growth of the bosonic string
partition function based on an integral over this subdomain may be
sufficient, because the measure is positive on all of moduli space, as a result
of the positive-definiteness of imaginary part of the period matrix and the
feasibility of expressing the rest of the measure as the squared modulus
of a holomorphic section of a line bundle over moduli space [12][13].

\noindent{\bf 2. Configurations of Isometric Circles and Bounds for the
Regularized
Integral}

There is a one-to-one correspondence between points in the fundamental domain
of the uniformizing Schottky group $\Gamma$ and the Riemann surface [14].  The
identification of points on the handle with points in the fundamental domain
can be achieved by identifying specific circles outside the isometric circles,
representing the boundaries of the removed disks in ${\hat {\Bbb C}}$, with
boundaries
of the bases of handles and joining the isometric circles at the middle of the
handle.  About each handle, there exists a closed curve which is geodesic.
The genus-independent cut-off on the length of closed geodesics can be
translated, therefore, to constraints on the sizes of the isometric circles.

For $g \ge 2$, the universal covering surface of the Riemann surface is the
Poincare disk or the upper-half plane with a hyperbolic metric, that can be
projected onto an intrinsic metric with constant curvature R = -1.
The restriction to R = -1 metrics defines a slice in the space of metrics,
which
is necessary in the reduction of the Polyakov path integral to a moduli
space integral [15], and it implies that the area increases linearly with the
genus.

Now consider the stereographic projection from ${\hat{\Bbb C}}$ to the sphere.
The Riemann
surfaces can be constructed directly by joining the isometric circles
to create handles.  Since the stereographic projection is genus-independent,
the dimensions of the sphere in the target space will remain unchanged as the
genus increases, but the dilation of the areas, and therefore the distances,
in the intrinsic metric on the sphere, with g handles, forces a decrease in
the size of the isometric circles, the spherical condition $\vert
\gamma_n\vert^{-2}\sim {1\over g}$ is required for consistency with the
cut-off on closed geodesic lengths.  Moreover, the relation,
$${{\vert \gamma_n\vert^{-2}}\over {\vert \xi_{1n}-\xi_{2n} \vert^2}}
{}~=~{{\vert K_n\vert}\over {\vert 1-K_n \vert^2}}
\eqno(7)$$
can be used to translate this condition to constraints on the multipliers and
fixed points.

As the relation (7) holds for transformations on the extended complex plane,
the
mapping of disks on ${\hat {\Bbb C}}$ to the sphere must be used before placing
restrictions
on the multipliers and fixed points.  If the isometric circles are confined
to a bounded region in ${\hat {\Bbb C}}$, the genus-dependence of $\vert
\gamma_n\vert^{-1}$
will be unaltered by the stereographic projection.  There are three categories
of isometric circles consistent with the condition $\vert \gamma_n\vert^{-2}
\sim {1\over g}$ in a bounded region in ${\hat {\Bbb C}}$.

$~~(i)~{{\epsilon_0}\over g} \le \vert K_n\vert \le {{\epsilon_0^\prime}\over
g}~~~~~~~~~~~~~
\delta_0\le \vert \xi_{1n} - \xi_{2n}\vert \le \delta_0
^{\prime}$
\vskip .05in
$~(ii)~{{\epsilon_0}\over {g^{1-2q}}}\le \vert K_n\vert \le
{{\epsilon_0^{\prime}}\over {g^{1-2q}}}~~~~~
{{\delta_0}\over {g^q}} \le \vert \xi_{1n}-\xi_{2n}\vert \le
{{\delta_0^{\prime}}\over {g^q}}~~~~~ 0 < q <{1\over 2}$
\vskip .05in
$(iii)~\epsilon_0 \le \vert K_n\vert \le \epsilon_0^{\prime}~~~~~~~~~~~~~
{{\delta_0}\over {\sqrt{g}}}\le \vert \xi_{1n}-\xi_{2n}\vert \le
{{\delta_0^{\prime}}\over {\sqrt{g}}}$

For the first category of configurations, lower bounds can be placed on the
primitive-element products and determinant factor in equation (2).
Since
$$\eqalign{\prod_\alpha~^\prime \vert 1 -
K_\alpha\vert^{-1}~&>~exp\left(-\sum_\alpha~^\prime \vert K_\alpha\vert\right)
\cr
\sum_\alpha~^\prime \vert K_\alpha\vert~&<~\sum_{n_l} {{\vert
\gamma_{n_l}\vert^{-2}}
\over {\left\vert\xi_{1n_l}+{{\delta_{n_l}}\over {\gamma_{n_l}}}\right\vert^2}}
+\sum_{n_l} {{\vert\gamma_{n_l}\vert^{-2}}\over {\left\vert
\xi_{2n_l}-{{\alpha_{n_l}}\over {\gamma_{n_l}}}\right\vert^2}}
+\sum_{V_{\tilde\alpha}=T_{n_l}V_\beta} {{\vert\gamma_{\tilde\alpha}\vert^{-2}}
\over {\left\vert \xi_{1\tilde\alpha}+{{\delta_{\tilde\alpha}}\over
{\gamma_{\tilde\alpha}}}\right\vert^2}}
\cr}
\eqno(8)$$
it suffices to estimate the growth of the bounds of these sums.
As $\vert \gamma_{n_l}\vert^{-2}~<~ {{\epsilon_0^\prime}\over
{\left[1-{{\epsilon_0^\prime}\over g}\right]^2}}{{\delta_0^{\prime 2}}\over
g}$,
the sum over the fundamental generators $\{I_{T_{n_l}^{\pm 1}}\}$ is less than
$2{{\epsilon_0^\prime}\over {\left[1-{{\epsilon_0^\prime}\over g}\right]^2}}
{{\delta_0^{\prime 2}}\over {\delta_0^2}}$.  The terms in the remaining sum can
be grouped according to the number of generators in $V_{\tilde\alpha}$.
When the product consists of two elements,
$V_{\tilde\alpha}=T_{n_{l_1}}T_{n_{l_2}}$, and
$$\vert\gamma_{\tilde\alpha}\vert^{-2}~=~\vert \gamma_{n_{l_1}}\vert^{-2}
\vert \gamma_{n_{l_2}}\vert^{-2} \left\vert {{\delta_{n_{l_1}}}\over
{\gamma_{n_{l_1}}}} + {{\alpha_{n_{l_2}}}\over
{\gamma_{n_{l_2}}}}\right\vert^{-2}
{}~\le~{{\epsilon_0^{\prime 2}}\over {\left[1-{{\epsilon_0^\prime}\over
g}\right]^4}} {{\delta_0^{\prime 4}}\over {g^2}} \left\vert {{\delta_{n_{l_1}}}
\over {\gamma_{n_{l_1}}}}+{{\alpha_{n_{l_2}}}\over {\gamma_{n_{l_2}}}}
\right\vert^{-2}
\eqno(9)$$

Now consider a dense packing of isometric circles spaced apart by a distance of
$O({1\over {\sqrt g}})$ in a finite region of the complex plane.
In a hexagonal arrangement of 2g circles, the circles can be indexed by the
level number l, defined by their distance from the origin, the number of
circles at level l is 6(l-1), and the number of levels $[{1\over 2}+{1\over 6}
{\sqrt{9+24 g}}]$.

The level numbers of $I_{T_{n_{l_1}}}$ and $I_{T_{n_{l_2}}^{-1}}$ can be chosen
to be $l$ and $l+l_0$ respectively.  Indexing the circles at these levels by
$j_1=0, 1, ..., 6(l-1)-1$ and $j_2=0,1,...,6(l+l_0-1)-1$, it follows that
$$\left\vert {{\delta_{T_{n_{l_1}}^{(j_1)}}}\over
{\gamma_{T_{n_{l_1}}^{(j_1)}}}} + {{\alpha_{T_{n_{l_2}}^{(j_2)}}}\over
{\gamma_{T_{n_{l_2}}^{(j_2)}}}}\right\vert^{-2}
\le \left[ d_{l+l_0}^2 +d_l^2-2d_ld_{l+l_0} cos \left({{i(j_1,j_2) \pi}\over
{3(l+l_0-1)}}\right)\right]^{-1}
\eqno(10)$$
where $d_l$ is the distance from  a level l circle to the center of the
configuration and $i(j_1,j_2)$ takes integer values in the range $[0,
6(l+l_0-1)-1]$.  Since the distance between the fixed points is greater than
$\delta_0$, there is no correlation between the distances
$d(I_{T_{n_{l_1,j_1}}}, I_{T_{n_{l_2,j_2}}^{-1}})$ and
$d(I_{T_{n_{l_1,j_1}}^{-1}}, I_{T_{n_{l_2,j_2}}})$.  A lower bound for the
latter distance can be obtained through the densest packing of $6(l+l_0-1)$
circles $\{I_{T_{n_{l_2}}}\}$ about $I_{T_{n_{l_1,j_1}}^{-1}}$.
As the average value of $[d(I_{T_{n_{l_1,j_1}}^{-1}},
\{I_{T_{n_{l_2,j_2}}}\}]^{-2}$ in this packing is about  ${g\over
{2\delta_0^2}}
{{ln(l+l_0-1)}\over {l+l_0-1}}$, the sum over the elements
$V_{\tilde\alpha}=T_{n_{l_1}}^{\pm 1}T_{n_{l_2}}^{\pm 1}$ in
$\sum_\alpha~^\prime \vert K_\alpha\vert$ is less than
$$\eqalign{\epsilon_0^{\prime 2} &\left[1-{{\epsilon_0^\prime}\over
g}\right]^{-4}
{{\delta_0^{\prime 4}}\over {\delta_0^4}}
\cr
&\cdot \sum_{l=2}^{[{1\over 2}+{1\over 6}{\sqrt{9+24g}}]} 18[l-1]
\biggl[1+{1\over {2l-2}}+{1\over 2}
\sum_{{l_0=-l+2}\atop {l_0\not= 0}}^{[{1\over 2}+{1\over 6}
{\sqrt{9+24g}}]-l} {{ln(l+l_0-1)}\over {l+l_0-1}} \left\vert {1\over {l_0}}
+{1\over {2l+l_0-2}}\right\vert \biggr]
\cr}
\eqno(11)$$
This bound increases linearly with respect to the genus, a property which
continues to hold for the higher-order terms, since summation over the elements
$V_{\tilde\alpha}=T_{n_{l_1}}^{\pm 1}...T_{n_{l_r}}^{\pm 1}$ introduces r-1
factors of the form $6[l^{T_{n_{l_s}}^{\pm 1}}-1]$ producing a term of order
$O(g^{r-1})$, while iteration of the formula (9) leads to a term of
order $O(g^{-(r-2)})$ associated with the factors $\vert
\gamma_{n_{l_s}}\vert^{-2}$.
As
$$\prod_\alpha~^\prime \prod_{p=1}^\infty \vert 1-K_\alpha^p\vert^{-48}~>~
\left(1-{{\epsilon_0^\prime}\over g}\right)^{{{48}\over
{{{\epsilon_0^\prime}\over g}\left(1-{{\epsilon_0^\prime}\over g}\right)}}
\sum_\alpha~^\prime \vert K_\alpha\vert}
\eqno(12)$$
the primitive-element products are bounded by an exponentially decreasing
function of the genus.
Similarly,
$$[det~Im~\tau]^{-13}~>~(2\pi)^{13 g}
\left[ln~g~-~ln~\epsilon_0~+~O(1)\right]^{-13g}
 \eqno(13)$$
since
$${{tr(Im~\tau)}\over g}~<~ {1\over {2\pi g}}\sum_{n=1}^g
\sum_\alpha~^{(n,n)}~{{\vert\gamma_\alpha\vert^{-2}\vert
\xi_{1n}-\xi_{2n}\vert^2}\over {\vert \xi_{1n}-V_\alpha\xi_{1n}\vert \vert
\xi_{2n}-V_\alpha\xi_{2n}\vert \left\vert \xi_{1n}+{{\delta_\alpha}\over
{\gamma_\alpha}}\right\vert \left\vert \xi_{2n}+{{\delta_\alpha}\over
{\gamma_\alpha}}\right\vert}}
\eqno(14)$$
and the dominant contribution to the second term arises from elements $T_{n_l}$
obeying equalities of the form $\vert \xi_{1n}-T_{n_l}\xi_{1n}\vert=O({1\over
{\sqrt g}})$, $\vert \xi_{2n}-T_{n_l}\xi_{1n}\vert =O(1)$,
$\left\vert \xi_{1n}+{{\delta_{n_l}}\over {\gamma_{n_l}}}\right\vert =O(1)$
and $\left\vert \xi_{2n}+{{\delta_{n_l}}\over {\gamma_{n_l}}}\right\vert
=0(1)$,
together with permutations of these relations.
   Integration
over the multipliers and fixed points in the above range leads to a lower bound
$$\eqalign{{ {\vert \xi_{11}^0 -\xi_{1g}^0\vert^2 \vert
\xi_{21}^0-\xi_{1g}^0\vert^2}
\over {\vert \xi_{11}^0-\xi_{21}^0\vert^2} } {{e^{-4\epsilon_0^{\prime}}}\over
2}&(2 \pi)^{16g-3} \left({1\over {\epsilon_0^2}}-{1\over {\epsilon_0^{\prime
2}}}\right)^g \left({2\over {\delta_0^2}}-{2\over {\delta_0^{\prime 2}}}
\right)^{g-1}\left(\delta_2^{\prime 2}-\delta_2^2 \right)^{g-2}
e^{-k(0) g}
\hfill
\cr
\cdot &~{{g^{2g}}\over {\left[ln g -ln \epsilon_0 + O(1)\right]^{13g}}}\cr}
\eqno(15)$$
where $\delta_2$ and $\delta_2^{\prime}$ represent the inner and outer radii
of an annulus containing the isometric circles and k(0) is a genus-independent
constant determined by the primitive-element products
$$k(0)~=~52~lim_{g\to \infty} {1\over g} \sum_\alpha~^\prime \vert
K_\alpha\vert
\eqno(16)$$
  The global SL(2,${\Bbb C}$) invariance of the Schottky uniformization has
been
used to fix the values of three points $\xi_{11},~\xi_{21}$ and $\xi_{1g}$.

For category (ii), a similar bound is obtained, except that the exponential
bound for the primitive-element products is $e^{-k(q)g}$ and the dominant
growth is given by
$${{g^{2(g-gq-q)}}\over {\left[(1-2q)ln g -ln \epsilon_0
+O(1)\right]^{13 g}}}$$

For the third category of isometric circles, the integral over the multipliers
of the group gives an exponential function of the genus, while the integral
over the fixed points is bounded below by a function which increases as
$c_1 c_2^g g^{g-1}$.  A lower bound can also be placed on the determinant
factor
$$\eqalign{(det~ Im~ \tau)^{-13}
>&(2 \pi)^{13g} \biggl[ ln \left({1\over {\epsilon_0}}\right)
\cr
&~~+{1\over g} \sum_{n=1}^g \sum_{\alpha}~^{(n,n)}
{{\vert \gamma_\alpha\vert^{-2} \vert \xi_{1n} - \xi_{2n} \vert^2}
\over {\vert \xi_{1n}+ {{\delta_{\alpha}}\over {\gamma_\alpha}}\vert
\vert \xi_{2n} -{{\delta_\alpha}\over {\gamma_\alpha}}\vert \vert
\xi_{1n}-V_\alpha
\xi_{1n} \vert \vert \xi_{2n}-V_\alpha \xi_{2n} \vert}}\biggr]^{-13 g}
\cr}
\eqno(17)$$
The second term can be regarded as the sum of a series involving fundamental
generators and their inverses and a series involving elements that are equal
to a product of two or more generators.  Finiteness of the term may be verified
through the convergence of each type of series.

\noindent{\bf Proposition 1}.  The remainder term in the lower bound for
$[det(Im~\tau)]^{-13}$ is finite for arbitrary genus.

\noindent{\bf Proof}.  As the minimum distance between neighbouring isometric
circles
in the bounded domain is ${ {\delta_0} \over {\sqrt{g}} }$, the configuration
involving the greatest number of circles in a region of fixed area would
involve
a hexagonal arrangement.  Assuming that $\xi_{1n}$ and $\xi_{2n}$ lie at the
center of the configuration, the isometric circles may be labelled as
$I_{T_{n_l}}$ and $I_{T_{n_l}^{-1}}$, where l refers to the level of
$I_{T_{n_l}}$ with respect to $I_{T_n}$.  As the number of circles at level l
is
6(l-1), the number of levels in a hexagonal configuration is $[{1\over 2} +
{1\over 6} \sqrt{9 + 24 g}]$, and since $\vert \xi_{2n}+{{\delta_{n_l}}\over
{\gamma_{n_l}}}\vert \ge (l-1) {{\delta_0}\over \sqrt{g}}$,
$$\eqalign{\sum_{n_l \ne n}& {{\vert \gamma_{n_l} \vert^{-2}
\vert \xi_{1n} -\xi_{2n}\vert^2}\over {\vert \xi_{1n}-T_{n_l} \xi_{1n}\vert
\vert \xi_{2n} -T_{n_l}\xi_{2n}\vert \left\vert \xi_{1n}+{{\delta_{n_l}}\over
{\gamma_{n_l}}}\right\vert \left \vert \xi_{2n} + {{\delta_{n_l}}\over
{\gamma_{n_l}}}\right\vert}}
\cr
+&
\sum_{n_l \ne n} {{\vert \gamma_{n_l}\vert^{-2} \vert\xi_{1n}-\xi_{2n}\vert^2}
\over {\vert \xi_{1n} - T_{n_l}^{-1} \xi_{1n} \vert \vert \xi_{2n} -
T_{n_l}^{-1} \xi_{2n}\vert \left \vert \xi_{1n} - {{\alpha_{n_l}}\over
{\gamma_{n_l}}}\right\vert \left \vert \xi_{2n} - {{\alpha_{n_l}}\over
{\gamma_{n_l}}}\right\vert}}
\cr
\le & 6{{\delta_0^{\prime 4}}\over {\delta_0^4}}
{{\epsilon_0^{\prime}}\over
{[1 - \epsilon_0^{\prime}]^2}} ~\sum_{l=2}^{2{{\delta_0^{\prime}}\over
{\delta_0}} \left[1 + {{\epsilon_0^{\prime {1\over 2}}}\over
{1 - \epsilon_0^{\prime}}}\right] }~~~[l-1]
+\Biggl[[6 {{\delta_0^{\prime 4}}\over {\delta _0^4}}
{{\epsilon_0^{\prime}}\over {[1- \epsilon_0^{\prime}]^2}}~
\cr
\cdot\sum_{2{{\delta_0^{\prime}}\over {\delta_0}}\left[1 +{{\epsilon_0^{\prime
{1\over 2}}}\over {1 - \epsilon_0^{\prime}}}\right] + 1}^{[{1\over 2}+{1\over
6} \sqrt{9+24 g}]}&
{{[l-1]}\over {\left[ [l-1] \left[l-1-{{\delta_0^{\prime}}\over {\delta_0}}
\right]\left[l-1-{{\delta_0^{\prime}}\over {\delta_0}}\left[1+{{2
\epsilon_0^{\prime {1\over 2}}}\over {1 - \epsilon_0^{\prime}}}\right]\right]
\left[l-1-2{{\delta_0^{\prime}}\over {\delta_0}}\left[1+
{{\epsilon_0^{\prime {1\over 2}}}\over {1 - \epsilon_0^{\prime}}}\right]\right]
\right]}}\Biggr]
\cr}
\eqno(18)$$
When $\xi_{2n}$ lies in the level $l^{\prime\prime}+1$, it is located at a
distance greater than ${l^{\prime\prime}}{{\delta_0}\over {\sqrt{g}}}$ from the
center of the configuration, and
a similar bound is valid with $[{1\over 2}+{1\over 6} \sqrt{9+24g}]$
replaced by $[{1\over 2}+{1\over 6} \sqrt{9+24 g}]+l^{\prime\prime}$.

The second type of series, involving a summation over elements
$V_{\tilde\alpha}=T_{n_l}V_{\beta}$, can be bounded even though the location of
the isometric circles $I_{V_{\tilde\alpha}}$ and $I_{V_{\tilde\alpha}^{-1}}$
may not be correlated.
The ratio of the inverse square of the radii of the circles
$I_{V_{\tilde\alpha}}$
and $I_{V_{\tilde\beta}}$ obeys the inequality
$$\left\vert {{\gamma_{\tilde\alpha}}\over {\gamma_\beta}}\right\vert^{-2}~=~
\vert \gamma_{n_l}\vert^{-2}\left\vert {{\delta_{n_l}}\over {\gamma_{n_l}}}
{}~+~{{\alpha_\beta}\over {\gamma_\beta}}\right\vert^{-2}~<~
{{\epsilon_0^{\prime }}\over {[1-\epsilon_0^{\prime}]^2}} {{\delta_0^{\prime
2}}\over g}\left\vert {{\delta_{n_l}}\over {\gamma_{n_l}}}+{{\alpha_\beta}
\over {\gamma_\beta}}\right\vert^{-2}
\eqno(19)$$
For elements that are products of two generators, or their inverses, it follows
that
$$\eqalign{~&{{\vert \gamma_{\tilde\alpha}\vert^{-2}\vert
\xi_{1n}-\xi_{2n}\vert^2}
\over {\vert \xi_{1n}-V_{\tilde\alpha} \xi_{1n}\vert \vert
\xi_{2n}-V_{\tilde\alpha}
\xi_{2n}\vert \left\vert \xi_{1n}+{{\delta_{\tilde\alpha}}\over
{\gamma_{\tilde\alpha}}}
\right\vert \left\vert \xi_{2n}+{{\delta_{\tilde\alpha}}\over
{\gamma_{\tilde\alpha}}}
\right\vert}}
\cr
&~~~~~~~~~~~~~~~~~~~~~~~~~~\le~
{{\epsilon_0^{\prime 2}}\over {[1-\epsilon_0^{\prime}]^4}}{{\delta_0^{\prime
6}}
\over {g^3}} {{\left\vert {{\delta_{n_l}}\over
{\gamma_{n_l}}}~+~{{\alpha_\beta}
\over {\gamma_\beta}}\right\vert^{-2}}\over {\vert
\xi_{1n}-V_{\tilde\alpha}\xi_{1n}
\vert \vert \xi_{2n}-V_{\tilde\alpha} \xi_{2n}\vert \left \vert \xi_{1n}+
{{\delta_{\tilde\alpha}}\over {\gamma_{\tilde\alpha}}}\right\vert \left\vert
\xi_{2n}+{{\delta_{\tilde\alpha}}\over
{\gamma_{\tilde\alpha}}}\right\vert}}\cr}
\eqno(20)$$
Using this inequality, an upper bound can be obtained for series involving
elements $V_{\tilde\alpha}$, which are products of two or more fundamental
generators.  The upper bounds for the higher-order terms have the same form
as the bound in equation (18), with coefficients ${{\delta_0^{\prime 2}}
\over {\delta_0^2}}\left[6{{\delta_0^{\prime 2}}\over {\delta_0^2}}
{{\epsilon_0^\prime}\over {(1-\epsilon_0^\prime)^2}}\right]^j$, multiplying
jth order sums over fractions involving the levels
\hfil\break
 $l^{T_{n_{l_1}}^{\pm 1}},...,
l^{T_{n_{l_j}}^{\pm 1}}$ having leading order behaviour $[l^{T_{n_{l_1}}^{\pm
1}} ]^{r_1}...[l^{T_{n_{l_j}}^{\pm 1}} ]^{r_j}$ where $\sum_{i=1}^j r_i~=~
-(j+2)$.  This suggests that the entire series containing $\xi_{1n}$ and
$\xi_{2n}$ is bounded by a geometric series that converges when
$6{{\delta_0^{\prime 2}}\over {\delta_0^2}}{{\epsilon_0^{\prime}}\over
{[1-\epsilon_0^{\prime}]^2}}<\Delta$, where $\Delta$ is determined by
equation (18).

Thus, the lower bound for $[det(Im~\tau)]^{-13}$ only decreases exponentially.
By evaluating the series $\sum_\alpha^{\prime}\vert K_\alpha\vert~=~
\sum_{\alpha}^{\prime} {{\vert \gamma_\alpha\vert^{-2}} \over
{\vert \xi_{1\alpha}+{{\delta_\alpha}\over {\gamma_\alpha}}\vert^2}}$ for the
hexagonal
configuration of isometric circles, it may be verified that the primitive-
element product $\prod_\alpha^{\prime} \vert 1-K_\alpha\vert^{-1}$ is also
bounded below by an exponentially decreasing function of the genus.  Combining
the lower bounds for the primitive-element products and determinant factor,
and the integrals over the multipliers and fixed points, a factorial increase
is
obtained for the entire integral of the measure (2) in the range for the
category (iii).

A procedure similar to the one outlined in this section can be used to set
upper bounds for the
integrals over each of the three ranges of values for $\vert K_n\vert$
and $\vert \xi_{1n} - \xi_{2n} \vert$, n=1,...,g [16].

\noindent{\bf 3. The Fundamental Region of the Symplectic Modular Group}

A rigorous bound can be found for the regularized partition function if the
integration region lies within the fundamental domain for the mapping class
group.  If the coordinates of the Siegel upper half plane are $(\tau_{mn})$,
where $det(Im~\tau)>0$, then the restrictions defining the fundamental region
of the symplectic modular group [17] are
$$\eqalign{(i)&~~-{1\over 2} \le Re~\tau_{mn} \le {1\over 2}
\cr
(ii)&~~\vert det(C \tau +D)\vert \ge 1~~~~~\left(\matrix{A&B \cr
                                                          C&D \cr}\right) \in
Sp(2g,{\Bbb Z})
\cr
(iii)&~~ Im~\tau [g_r]\equiv g_r^T (Im~\tau) g_r \ge Im~\tau [e_r] \cr
&~~where ~e_r = (0,...,0,1,0,...0) ~and ~g_{rr},...,g_{rg} ~are ~relatively
{}~prime
\cr
&~~(Im~\tau)_{1r} \ge 0
\cr}
\eqno (21)$$
While these conditions can be translated to restrictions on the Schottky
coordinates by inverting the formula (3) for the period matrix elements, it is
simpler to verify that refined limits would lead  to the matrix inequalities
being satisfied.

An analysis of the real part of the period matrix reveals that the arguments of
the fixed point ratios tend to cancel, so that the dominant contribution to
$Re ~\tau_{nn}$ shall be the argument of $K_n$.  A shift in the argument of
$K_n$ by $2 \pi$ is equivalent to a shift in $Re ~\tau_{nn}$.  Since the
transformation
$$\tau ~\to~ \tau~+~\left(\matrix{0&~~~~~~~~~~&\cr
                          &.~~~~~~~~~&\cr
                          &~.~~~~~~~~&\cr
                          &~~.~~~~~~~&\cr
                          &~~~0~~~~~~&\cr
                          &~~~~1~~~~~&\cr
                          &~~~~~0~~~~&\cr
                          &~~~~~~.~~~&\cr
                          &~~~~~~~.~~&\cr
                          &~~~~~~~~.~&\cr
                          &~~~~~~~~~0&\cr}\right)
\eqno(22)$$
is symplectic, generalizing the shift $\tau \to \tau +1$ at one loop, a
translation of the argument of $K_n$ by $2\pi$ moves the point in parameter
space to a different fundamental region.  Since the range for arg $K_n$ will
not be reduced significantly by the fixed point ratios, except possibly
for the third category of isometric circles, it will be sufficient to
integrate,
in general, over a interval approximately equal to $[- \pi, \pi].$

To verify the inequality $\vert det(C \tau +D)\vert \ge 1$, representing the
second condition defining the fundamental region, it is preferable to
split the determinant
$$\vert det(C \tau +D)\vert~=~\vert det(Im~\tau)\vert \vert det
(C-iC(Re~\tau)(Im~\tau)^{-1}-iD(Im~\tau)^{-1})\vert
\eqno(23)$$
For the first two types of configurations, the diagonal entries of $Im~\tau$
grow as $O(ln g)$ and the Minkowski inequality [18] for positive-definite
matrices
implies
$$det(Im~\tau) \ge {{(Im~\tau)_{11}...(Im~\tau)_{gg}}\over {\Gamma({g\over 2}
+1)]^2}} \left({\pi \over 4}\right)^g \left({2\over 3}\right)^{g(g-1)}
\sim O\left({{ln ~g}\over g}\right)^g \left({e\over 4}\right)^g
\left({2\over 3}\right)^{g(g-1)}
\eqno(24)$$
As the determinant is also equal to the volume of the parallelepiped spanned
by the basis vectors ${\underline v}_1$,..., ${\underline v}_g$ representing
the rows of the matrix,
$\vert {\underline v}_1\vert~...~\vert{\underline v}_g\vert
sin\theta_{12}~...~sin \theta_{1...g}$,
where $\theta_{1...n}$ is the angle from ${\underline v}_n$ to the hyperplane
spanned
by ${\underline v}_1$, ..., ${\underline v}_{n-1}$, and since
\hfil\break
$sin \theta_{1...n}=O(1)$
when the off-diagonal entries decrease as $O({1\over{\sqrt{g}}})$, it follows
that
\hfil\break
$det (Im~\tau) \ge  d(q)^g O(ln g)^g$ if the off-diagonal elements have a
magnitude decreasing with the
genus at that rate.

To determine the effect of the off-diagonal elements on the lower bound for the
determinant, the following inequalities for eigenvalues of hermitian matrices
[19] are relevant.  If $\lambda_1 \le \lambda_2 \le ... \le \lambda_g$ are the
eigenvalues, $\sigma_k^{(n)}$ denotes the sum of the absolute values of
off-diagonal elements of the kth row of any principal submatrix of order n and
$h_k^{(n)}$ represents the kth diagonal element of the prinicipal submatrix,
then
$$min_k [h_k^{(g-n+1)} - \sigma_k^{(g-n+1)}] \le \lambda_n \max_k [h_k^{(n)}
+\sigma_k^{(n)}]
\eqno(25)$$
In particular, $\lambda_1 \le min_n (Im~\tau)_{nn}$ and $\lambda_g \ge
max_n(Im~\tau)_{nn}$.  Given the period matrix with imaginary part having
diagonal entries ${1\over {2\pi}}ln~g+O(1)$ and off-diagonal entries
$\beta_{mn}$, the following inequalities hold
$$\eqalign{\lambda_g~&\ge~O(ln~g)
\cr
{1\over {2\pi}}ln~g+O(\vert (g-2)\cdot \beta_{mn}\vert)~&\ge~\lambda_{g-1}
{}~\ge~ {1\over {2\pi}}ln~g~-~O(\vert \beta_{mn}\vert)
\cr
{1\over {2\pi}}ln~g+O(\vert (g-3)\cdot \beta_{mn}\vert)~&\ge~\lambda_{g-2}
{}~\ge~ {1\over {2\pi}}ln~g~-~O(2\vert \beta_{mn}\vert)
\cr
&...
\cr
{1\over {2\pi}}ln~g~&\ge~\lambda_1~\ge {1\over {2\pi}}ln~g~-~O(\vert (g-1)\cdot
\beta_{mn}\vert)
\cr}
\eqno(26)$$

To obtain a bound on the lowest eigenvalue of the type $\lambda_1 \ge O(\vert
\beta_{mn}\vert)$, it is required that $\sigma_k^{(j)} \le
{1\over {2\pi}}ln~g-O(\vert\beta_{mn}\vert)$.  This would occur if the
off-diagonal elements
decrease as $O\left({{ln~g}\over g}\right)$.  Under these conditions,  the
determinant is greater than $\left({1\over {2\pi}}ln~g\right)^{O(ln~g)}$ is
obtained.
When the off-diagonal entries are O(1), a property which holds for the three
categories of isometric circles mentioned above, then a lower bound for the
determinant can be obtained using the identity $det\left(\matrix{A&B&
                                                              \cr
                                                                C&D&\cr}\right)
      =detA~det(D-CA^{-1}B)$ and verifying that the
determinant does decrease when the diagonal entries are reduced and
the off- diagonal entries are increased.  If the diagonal elements are
chosen to be equal intially, and denoted by $\alpha$, and $\beta_1 <
\beta_{mn} < \beta_2$, the bound obtained from iteration of the
identity is
$$\eqalign{\alpha^{2-g}&
(\alpha-\beta_2)^{g-1}(\alpha+\beta_1)^{3-g}
(\alpha+\beta_1-\beta_2)^{g-2} \cr
&(\alpha+2\beta_1)^{4-g}(\alpha+2\beta_1-\beta_2)^{g-3}...
(\alpha+(g-3)\beta_1)^{-1} \cr
&(\alpha+(g-3)\beta_1-\beta_2)^2
(\alpha+(g-2)\beta_1-\beta_2)(\alpha+(g-1)\beta_1)
\cr &>(\alpha-\beta_2)^{g-1} (\alpha+(g-1)\beta_1)
e^{-{{(\beta_2-\beta_1)}\over {\beta_1}} g~ln~g}~~~~~as~g~\to~\infty
\cr} \eqno(27)$$
By overlapping the intervals $\{[\beta_1,
\beta_2]\vert~\vert \beta_2-\beta_1\vert={1\over ln~g}\}$ until
$\beta_1$ reaches ${{ln~g}\over g}$, where the theorem about
eigenvalues of matrices and their principal submatrices can be used,
 monotonic growth of the bound with respect to the genus is seen to hold over
the interval 0 to $\beta_{mn}=\vert O(1)\vert$.

When the diagonal elements of $Im~\tau$ are $O(ln~g)$, the proportion of
matrices with $[det(Im~\tau)]=O(1)$ decreases rapidly with the genus.  First,
it
may be noted that the determinant vanishes when the diagonal elements equal
${1\over {2\pi}}ln~g$ and the off-diagonal entries all equal
$-{1\over {2\pi}}{{ln~g}\over {g-1}}$ and is O(1) only if $\beta_{mn}=-{1\over
{2\pi}} {{ln~g}\over {g-1}}+O\left({{(2\pi)^{g-1}}\over
{(g-1)(ln~g)^{g-1}}}\right)$.  If the off-diagonal elements are assumed to be
O(1), then the expression for the determinant defined earlier, $\vert
{\underline v}_1\vert~...~\vert {\underline v}_g\vert sin \theta_{12}~...~sin
\theta_{1...g}$, will be O(1) only if
$sin \theta_{12}~...~sin \theta_{1...g}=O\left({1\over {(ln~g)^g}}\right)$.
The analysis may be simplified if it is assumed that the off-diagonal entries
are set equal to $\pm 1$.  Let $n_2,...,n_g$ denote the number of negative
entries in the vector ${\underline v}_2,..., {\underline v}_g$ respectively.
Then
$$n_2~\simeq~{1\over 2} g~sin^2 \theta_{12}~~~...~~~n_g~\simeq~{1\over 2} g~
sin^2 \theta_{1...g}
\eqno(28)$$
As there are ${{g(g-1)}\over 2}$ independent off-diagonal entries in the
symmetric matrix $Im~\tau$, the number of positive entries in the upper
triangular part of the matrix can be denoted by ${{g(g-1)}\over 4} +m$, while
the number of negative entries would be ${{g(g-1)}\over 4}-m$.
Thus,
$$m\simeq {g\over 4} [~(g-1)~-~sin^2 \theta_{12}~-~...~-~sin^2 \theta_{1...g}]
\eqno(29)$$
It would appear that the product constraint could be satisfied when all but one
of the angles equals ${\pi \over 2}$.  However, orthogonality of (g-1) vectors
${\underline v}_n$ does not hold when the components of the vectors are
$\pm 1$.  Instead, the signs can be chosen so that only $O({\sqrt g})$ vectors
               are orthogonal.  From the constraint on the product
$sin\theta_{12}~...~sin \theta_{1...g}$, it follows that the minimum value of m
should take the form
$$\eqalign{m_{min}~&\simeq~{g\over 4}(g-1-O({\sqrt g}))
[1-f_1(g)...f_{g-1-O({\sqrt g})} (g)]
\cr
\prod_{i=1}^{g-1-O({\sqrt g})}~f_i(g)~&=~O\left({1\over {(ln~g)^g}}\right)~~~~
        f_i(g)~\to~0~~~i=1,..., g-1-O({\sqrt g})~~as~g\to\infty
\cr}
\eqno(30)$$
For large g, it may be assumed that $m \ge {{g(g-1)}\over 8}$.  To calculate
the proportion of matrices with m satisfying this inequality, one may note that
the signs of the off-diagonal entries of $Im~\tau$ are determined by the
arguments of the fixed-point terms and the parameters $\gamma_\alpha^{-2}$,
which will be randomly distributed in the range $[-\pi, \pi]$ for a large
configuration of isometric circles.  Thus, the probability of a positive or
negative entry will be approximately equal ${1\over 2}$, and the computation of
the probability becomes identical to the
statistical mechanical problem concerning the alignment of spins in a lattice.
The number of matrices N(m) with a given value of m is $N(0)e^{{-4m^2}\over
{g(g-1)}}$, and the fraction of matrices with $m \ge {{g(g-1)}\over 8}$
is less than ${{g(g-1)}\over 8} e^{-{{g(g-1)}\over {16}}}$.

It can be concluded, therefore, that for almost all of the configurations of
isometric circles in categories (i) and (ii), $\vert det(Im~\tau)\vert$ will be
a rapidly increasing function of the genus.
   When $det C\ne 0$,
$\vert \det(C-iC(Re~\tau)(Im~\tau)^{-1}-iD(Im~\tau)^{-1})\vert$ will be bounded
below by a constant, because C would have integer entries, whereas the
remaining matrices would have rapidly decreasing entries after diagonalization.
When C=0, the expression (23) becomes $\vert det D \vert \ge 1$ as
$det ~D\ne 0$ and D has integer entries.  The determinant condition defining
the
fundamental region will therefore be satisfied.

For the third category of isometric circles, the diagonal entries of $Im~\tau$
no longer increase logarithmically with the genus, but the determinant
inequality will still be valid if $\epsilon_0^{\prime}<< 1$.  In particular,
it can be deduced from the general expression (27) that the determinant will
be a monotonically increasing function of the genus
when
$$\epsilon_0^{\prime}< exp[-B_0(max_n s_{nn} + max_{m\not= n} s_{mn})]
\eqno(31)$$
where $s_{mn}$ is an upper bound for $\sum_\alpha~^{(m,n)}~ln\left\vert
{{\xi_{1m}-V_\alpha \xi_{2n}}\over {\xi_{1m}-V_\alpha \xi_{1n}}}
{{\xi_{2m}-V_\alpha \xi_{1n}}\over {\xi_{2m}-V_\alpha \xi_{2n}}}\right\vert$
and $B_0$ is an appropriate constant.

The third set of conditions defining the fundamental region comprise an
infinite set of inequalities for the period matrix inequalities.  In the
following proposition, it is shown that these inequalities generally reduce
to a special class of inequalities, which can be explicitly analyzed.

\noindent{\bf Proposition 2}. For sufficiently large genus, or for a
sufficiently
small upper limit for the magnitudes of the multipliers, the conditions
$g_r^T(Im~\tau) g_r \ge e_r^T(Im~\tau) e_r$ always hold when $(Im~\tau)_{ss}
\ge (Im~\tau)_{rr}$ for $s \ge r$.

\noindent{\bf Proof}. When only one entry between $g_{rr}$ and $g_{rg}$ is
non-zero,
the condition $Im~\tau[g_r] \ge Im~\tau[e_r]$ implies that $g_{rs} (Im~
\tau)_{ss} \ge (Im~\tau)_{rr}$ ,  for each s=r,...,g, which is valid when
$(Im~\tau)_{ss} \ge (Im~\tau)_{rr}$.

When $\vert K_n\vert = O(ln g)$ and the contribution from the multipliers
dominates the contribution from the fixed points, so that the sequentially
ordered eigenvalues of
$Im~\tau$ are $\lambda_n =O(ln g)$ for all n=1,...,g, the inequality
${{(Im~\tau)_{gg}}\over {\lambda_1}} < C_0$ for some constant $C_0$ is
satisfied.
If $g_r^Tg_r >C_0$, then $g_r^T(Im~\tau)g_r \ge \lambda_1 g_r^T g_r > (Im~
\tau)_{gg} \ge (Im~\tau)_{rr}$.
If $g_r^T g_r \le C_0$, then at most $C_0$ $g_{rn}$ are non-zero, including at
least one $g_{rs}, s\ge r$.  When all of the entries are equal to $\pm 1$, the
following inequalities will then hold, after appropriate relabelling of the
indices.
$$\eqalign{(Im~\tau)_{11}+...+(Im~\tau)_{C_0-1, C_0-1}+(Im~\tau)_{ss} &\ge
{{C_0-1}\over {2 \pi}}[(1-2q) ln~ g-ln \epsilon_0^{\prime}-max_n s_{nn}]
\cr
&~~~~~~~~
 +(Im~ \tau)_{rr},
\cr
&...~~~~~~~~~~~~~~s\ge r
\cr
\pm 2 (Im~\tau)_{12}\pm...\pm 2(Im~\tau)_{C_0-1,s} &\ge -{{C_0(C_0-1)}\over {2
\pi}}
max_{m\ne n} s_{mn}\cr}
\eqno(32)$$
The inequality
$g_r^T (Im~\tau) g_r \ge e_r^T (Im~\tau) e_r$ will then be
satisfied when
\hfil\break
$g \ge \epsilon_0^{\prime {1\over {1 - 2q}}} exp[{C_0\over {1-2q}} max_{m\ne n}
s_{mn} +
max_n s_{nn}]$.  Moreover, the derivative of the quadratic form
$g_r^T(Im~\tau) g_r$ with respect to $g_{rn}$ is non-negative when $g_r^T g_r
\le C_0$ and when the genus is greater than the lower bound just given.
Thus, for higher genus, the third set of reduction conditions is equivalent
to $(Im~\tau)_{ss} \ge (Im~\tau)_{rr}, s\ge r$.

The multipliers for isometric circles in the third category have absolute value
\hfil\break
$\vert K_n\vert = O(1)$, so that the diagonal entries of $Im~\tau$ in the
inequalities
$g_r^T (Im~\tau) g_r \ge e_r^T (Im~\tau) e_r$ are no longer
rapidly increasing functions of the genus.  However, the eigenvalues are
still positive and of the same order, so that the lowest eigenvalue satisfies
the inequality ${{(Im~\tau)_{gg}}\over {\lambda_1}} < C_0$ for some constant
$C_0$.
When $g_r^T g_r > C_0$, then $Im~\tau [g_r] \ge Im~\tau [e_r]$ immediately
follows
from $(Im~\tau)_{ss} \ge (Im~\tau)_{rr}$, $s \ge r$.  When $g_r^T g_r \le C_0$,
$(Im~\tau)_{ss} \ge (Im~\tau)_{rr}, s \ge r$ implies that the reduction
conditions hold if
$$\epsilon_0^{\prime} \le  exp[-C_0(max_n s_{nn} + max_{m\ne n} s_{mn})]
\eqno(33)$$

The inequalities $(Im~\tau)_{ss} \ge (Im~\tau)_{rr}$, $s \ge r$, lead to
constraints of the form ${{\vert K_1\vert}\over {\rho_1}} \ge {{\vert K_2\vert}
\over {\rho_2}} \ge ... \ge  {{\vert K_g\vert}\over {\rho_g}}$
where
$$\rho_n~=~\prod_\alpha~^{(n,n)}\left\vert{{\xi_{1n}~-~V_\alpha \xi_{2n}}\over
{\xi_{1n}~-~V_\alpha \xi_{1n}}}{{\xi_{2n}~-~V_\alpha \xi_{1n}}\over
{\xi_{2n}~-~V_\alpha \xi_{2n}}}\right\vert
\eqno(34)$$
Denoting  $\rho^{-1}= max \{\rho_1,...,\rho_g\}$, the factor
$\left({1\over {\epsilon_0^2}}~-~{1\over {\epsilon_0^{\prime 2}}}\right)^g$
in equation (15)
is replaced by
${1\over {\rho_1^2...\rho_g^2 g!}}\left({1\over {\rho^{-2} \epsilon_0^2}}
{}~-~{1\over {\rho_1^2 \epsilon_0^{\prime 2}}}\right)^g$.

For configurations of isometric circles in category (iii), both terms
involving the multipliers and fixed points in $(Im~\tau)_{nn}$ are of order
O(1) so that the inequalities $(Im~\tau)_{ss} \ge (Im~\tau)_{rr}, s \ge r$
may be satisfied by imposing constraints on different sets of parameters, other
than the values of $\vert K_n\vert$, n=1,...,g.
It has been shown that restriction of the arguments of $\xi_{1r}-\xi_{2r}$
and $K_r$ for isometric circles $I_{T_r}$ and $I_{T_r^{-1}}$ in a local
neighbourhood
consisting of J circles can be sufficient to obtain a sequential ordering of
$\{(Im~\tau)_{rr}\}$ for those circles when J is bounded [16].  This local
ordering can be
repeated for the entire configuration, reducing the integral over the
fixed points by an exponential function of the genus.  A representative
generator associated with each neighbourhood can then be chosen, and the
ordering of the $\{(Im~\tau)_{r_N r_N}\}, N=1,...,{g\over J}$
may be achieved either through constraints on the absolute values of the
multipliers $K_{r_N}$, the distances $\vert \xi_{1r_N}-\xi_{2r_N}\vert$
or both simultaneously.  A reduction of the combined integral over the
multipliers and fixed points by a factor of order $O({g\over J})!$
produces a dependence on the genus of ${{c_1 c_2^g c_3^{g\over J}}\over
{[det(Im~\tau)]^{13 g}e^{k({1\over 2}) g}}} g^{g\left(1-{1\over J}\right)}$.

To satisfy the inequality $(Im~\tau)_{1n} \ge 0$, suppose that the
configuration
of isometric circles has the property that $I_{T_2^{\pm 1}},...,I_{T_g^{\pm
1}}$
are located in a region of the extended complex plane, separated from
$I_{T_1^{\pm 1}}$, and that the distances from the fixed points $\xi_{11}$
and $\xi_{21}$ to the group of isometric circles is much greater than the
radii of these circles.  The restriction on $(Im~\tau)_{1n}$
implies that
$$\eqalign{\prod_\alpha~^{(1,n)} &\left\vert{{\xi_{11}~-~V_\alpha
\xi_{2n}}\over
{\xi_{11}~-~V_\alpha \xi_{1n}}} {{\xi_{21}~-~V_\alpha \xi_{1n}}\over {\xi_{21}
{}~-~V_\alpha \xi_{2n}}}\right\vert\cr
=&\prod_\alpha~^{(1,n)}\left\vert
1+{{(\xi_{21}~-~\xi_{11})(\xi_{1n}~-~\xi_{2n})\gamma_\alpha^{-2}}\over
{(\xi_{1n}~+~{{\delta_\alpha}\over {\gamma_\alpha}})
(\xi_{2n}~+~{{\delta_\alpha}\over {\gamma_\alpha}}) (\xi_{11}~-~V_\alpha
\xi_{1n})(\xi_{21}~-~V_\alpha \xi_{2n})}}\right\vert \cr
\ge& ~1}
\eqno(35)$$
Using the freedom to specify the parameters of the generators $T_n$, the
positions of the fixed points $\xi_{11}$ and $\xi_{21}$ and the arguments of
the parameters $\xi_{1n}~-~\xi_{2n}$ and $\gamma_n$ can be chosen so that the
remainder term
in the product in (35) has a phase between $-{\pi \over 2}$ and ${\pi \over
2}$,
giving rise to a fixed-point fraction with absolute value greater than one.
As the terms with $V_\alpha=T_m^{\pm 1}, m\ne 1,n$ provide the dominant
contribution to the product (35), it follows that the period matrix would
satisfy the reduction condition for these configurations.  The constraints
on the arguments reduce the integral over the selected regions in the
Schottky group parameter space by an exponential function of the genus.

Having obtained the contributions of the three categories of isometric
circles to the integral, it might appear that the continuous parameter q,
defining the dependence of the multipliers and fixed points on the genus
for the intermediate configurations, could also lead to an infinity.
However, it has been demonstrated that the requirement of non-overlapping
of the ranges ${{\epsilon_0}\over {g^{1-2q}}} \le \vert K_n\vert
\le {{\epsilon_0^{\prime}}\over {g^{1-2q}}}$ forces a selection of
approximately ${{ln~g}\over {ln \left({{\epsilon_0^{\prime}}\over
{\epsilon_0}}\right)}}$ evenly spaced values in the interval $[0,{1\over 2})$.
Summing over the allowed values of q gives a lower bound
$${{O(g^g)}\over {[ln g -ln \epsilon_0 +...]^{13 g}}}
\left[{1\over {1~-~e^{-2(g+1)ln ({{\epsilon_0^\prime}\over
{\epsilon_0}})}}}\right]
\eqno(36)$$

\vfill
\eject
\noindent{\bf 4. Other Configurations of Isometric Circles}

There exist other configurations of isometric circles, besides the three
categories
listed above,  consistent with the condition $\vert \gamma_n\vert^{-2}\sim
{1\over g}$, but their effect on the bounds for regularized partition function
is
not significant at large genus.  By considering the stereographic projection of
the extended
complex plane onto the sphere, it can be shown that a disk of area
$O(g^{2r-1})$ is projected onto a disk of area $O({1\over g})$ at a distance
$O(g^r), r>0$, from the origin.  The distances between the fixed points
in this configuration are $O(g^{r-q}), 0 \le q\le {1\over 2}$, while the
integral over $\vert \xi_{2n}\vert$ grows as $O(g^{2r})$.
By equation (7), $\vert  K_n\vert~=~O(g^{2q-1})$
so that integration over the multipliers and fixed points and multiplication
by the determinant factor $[det(Im~\tau)]^{-13}$ gives a bound
$$C(q)^g{{O(g^{(1-2q)g-2(r+q)})}\over {[(1-2q) ln g - ln \epsilon_0
+...]^{13g}}}
\eqno(37) $$
for $0 \le q  < {1\over 2}$
and
$$C({1\over 2})^g O(g^{g-2r-1})
\eqno(38)$$
for $q = {1\over 2}$.  These bounds have already included the reduction of the
integration range because of the conditions
defining the fundamental region of the group.
Although the range of values of the power r lie between 0 and $\infty$,
non-overlapping of the intervals $\delta_0 g^{r-q} \le \vert \xi_{1n} -
\xi_{2n}\vert \le \delta_0^{\prime} g^{r-q}$ requires that only an
evenly spaced discrete set of values of r, $N {{ln ({{\delta_0^{\prime}}
\over {\delta_0}})}\over {ln g}}, N = 0,..., \infty$ be chosen to
contribute to the bound for the regularized partition function.
Adding all of these contributions gives
$$C(q)^g {{O(g^{2g-2qg-2q}})\over {[(1-2q) ln g -ln \epsilon_0~+~...]^{13 g}}}
\left[{1\over {1~-~{{\delta_0^2}\over {\delta_0^{\prime 2}}} }}\right]
\eqno(39)$$
for $0 \le q < {1\over 2}$
and
$$C({1\over 2})^g O(g^{g-1})\left[{1\over {1~-~{{\delta_0^2}\over
{\delta_0^{\prime 2}}} }}\right]
\eqno(40)$$
for $q = {1\over 2}$.
The growth of the estimates with respect to g, associated with the three types
of configurations considered in the previous section, is essentially unchanged.

A larger contribution to the partition function arises from
configurations of isometric circles satisfying the condition
$\vert\gamma_n\vert^{-2} \ge O({1\over g})$.  These configurations are allowed,
because the cut-off condition on the lengths of closed geodesics does not
exclude surfaces with some of the handles having a thickness increasing
with the genus and larger than the genus-independent lower bound in the
intrinsic metric.  This leads to the inclusion of other categories of isometric
circles
$$\eqalign{{{\epsilon_0}\over {g^{1-2q}}}& \le \vert K_n\vert \le
{{\epsilon_0^\prime}\over {g^{1-2q}}}~~~~~\delta_0 \le
\vert\xi_{1n}-\xi_{2n}\vert\le \delta_0^\prime~~~~~~~~0<q<{1\over 2}
\cr
{{\epsilon_0}\over {g^{1-2q^\prime}}} &\le \vert K_n\vert \le
{{\epsilon_0^\prime}\over {g^{1-2q^\prime}}}~~~~~{{\delta_0}\over {g^q}}
\le \vert \xi_{1n}-\xi_{2n}\vert \le {{\delta_0^\prime}\over {g^q}}~~~~~~~~
0<q<q^\prime<{1\over 2}
\cr}
\eqno(41)$$

There will be a reduction in the integrals over the multipliers and fixed
points
in these categories, because the limits have a different genus-dependence.
The lower bounds for the three types of configurations considered in $\S 2$
and $\S 3$, together with the bounds associated with the additional
categories (41) can be combined as the genus-dependence of the magnitudes
of the multipliers and fixed-point distances can be assigned to different
subsets of the 2g isometric circles simultaneously.  To obtain a bound
corresponding to the inclusion of all of the allowed configurations, it is
useful to establish the form of the combinatorial factor [16] multiplying
the integral.  This factor depends on the distinction between different classes
of isometric circles, and in particular, the subdivision of the general
categories according to the different values of q and $q^\prime$.  Since
there are approximately ${1\over 2}{{(ln~g)^2}\over
{\left(ln({{\epsilon_0^\prime}\over {\epsilon_0}})\right)^2}}$ pairs $(q,
q^\prime)$, the number of groupings of multipliers and fixed points is
$${{\left(g+{1\over 2}{{(ln~g)^2}\over {\left(ln({{\epsilon_0^\prime}\over
{\epsilon_0}})\right)^2}}-1\right)!}
\over {g! \left({1\over 2} {{(ln~g)^2}\over {\left(ln({{\epsilon_0^\prime}\over
{\epsilon_0}})\right)^2}}\right)!}}
\simeq g\left({{2g \left(ln({{\epsilon_0^\prime}\over {\epsilon_0}})\right)^2}
\over {(ln~g)^2}}+1\right)^{{1\over 2} {{(ln~g)^2}\over
{\left(ln({{\epsilon_0^\prime}\over {\epsilon_0}})\right)^2}}}
\eqno(42)$$
While this factor increases the lower bounds found in $\S 3$, it is not
sufficient to obtain a factorial growth with respect to the genus, as
$g^{(ln~g)^2}$ does not compensate for division by the the logarithmic term
$(ln~g)^{13 g}$ in
categories (i) and (ii) or the factorial term $({g\over J})!$ in category
(iii).

The sum of the contributions of the different configurations of isometric
circles to the regularized partition function may be compared with the
lower bound obtained using the combinatorics of graph theory [1].  Viewing the
Riemann surfaces of finite genus as thickened trivalent graphs with branches
at the final level intertwined, the number of different pairings of the
branches increases at a factorial rate with respect to the genus [20].
While the number of annuli in Teichmuller space associated with the thickening
of each of the graphs grows at a factorial rate, the total contribution
will be reduced by intersections of the annuli and the condition of integration
over a single fundamental region of the symplectic modular group.  This may
require a study of the modular transformations which would map a Riemann
surface, constructed using spheres with 3 punctures (in a manner similar to
the Schottky uniformization) [21], to another surface built by joining the
spheres
at a set of interchanged punctures.

While there is agreement with the results of the graph theory argument
regarding
the rapid growth of the regularized partition function with respect to the
genus, a different conclusion is reached regarding Borel summability of the
resulting perturbation series.  Unless there are additional contributions to
the partition function associated with special configurations of isometric
circles, an exact factorial increase is not obtained and the dividing factors
$(ln~g)^{13 g}$ and $({g\over J})!$ cause the Borel-transformed series to be
convergent.  The lower bounds given in $\S 3$ are relevant, because it may be
shown that upper bounds can be found with the similar genus-dependence [16].
Moreover, these results suggest the problem of physically interpreting
perturbation series of the type $\sum_g {{g!}\over {(ln~g)^g}}$ and $\sum_g
({g\over k})!$.
Finally, whereas the combinatorics of graph theory appear to lead to a
large-order factorial growth,
independently of the field  or string content of the theory [22], the methods
used here may reveal more immediately the higher degree of finiteness of
superstring theories [23].
\vfill
\eject
\noindent{\bf 5. Growth of the Regularized Partition Function and the Domain of
String Perturbation Theory}

The calculation of amplitudes of string scattering processes has led to a
definition of the domain of string perturbation theory.  In particular, a
study of the scattering of four tachyons associated with a class of surfaces
uniformized by an infinitely-generated group of Schottky type suggested that
the sum over histories should only include a special category of effectively
closed manifolds [24].  This class would contain closed surfaces of finite
genus
and a set of infinite-genus surfaces which can be uniformized by a group of
Schottky type and belong to $O_G$ by classification theory [25].  This property
is consistent with a finite-size interaction region, which would exclude
surfaces of infinite extent in the target space, with ideal boundaries of
positive linear measure being observed as additional string states.
Restricting the domain of string perturbation theory to effectively closed
surfaces would appear to be required for consistency of the perturbative
expansion of the S-matrix.

The allowed configurations of isometric circles have been selected to be
consistent with a genus-independent cut-off on the length of closed geodesics
in the intrinsic metric on the Riemann surface [26].  The growth of the bounds
for the integrals over each range of values of $\vert K_n\vert$ and
$\vert\xi_{1n} -\xi_{2n}\vert$, n=1,...,g, reflects the geometrical
characteristics of the string worldsheet.  The class of effectively
closed surfaces with handles of thickness diminishing at a rate of
$\left({1\over {n^q}}\right), q>{1\over 2}$ [24], would be excluded by the
cut-off.  It might appear that configurations of isometric circles {\it on the
extended complex plane} with the square of the radii $\vert \gamma_n\vert^{-2}$
decreasing as $O({1\over n})$  would be consistent with a genus-independent
cut-off on the length of closed geodesics.  However, the latter condition
has been translated in the Schottky parametrization to the {\it sphere of
projection}.  Since the circles of area $O({1\over n})$ on the complex plane
are non-overlapping, they would occupy a region of infinite area.  A conformal
transformation to a bounded region in the complex plane, would decrease
the dependence of the areas to $\left({1\over {n^{2q}}}\right), q>{1\over 2}$.
Even in the intrinsic metric, which expands areas by a factor of g, the
handles would still be accumulating to a point.  This example confirms the
theorem proven earlier [24] relating surfaces that can be uniformized by groups
of Schottky type and those in the class $O_G$.
  For the manifolds considered in this paper, the length of the closed
geodesics are bounded below and the surfaces of infinite genus could be
expanded to be manifolds of infinite extent with boundary.  Moreover,  the
equivalent
condition on the radii of the isometric circles $\vert \gamma_n\vert^{-2}
\sim {1\over g}$ implies that the size would vanish in the infinite-genus
limit.  These particular surfaces of infinite genus, therefore, could not
be uniformized by a Schottky group and belong to $O_G$.
While these manifolds represent the geometrical limit of surfaces which lie
within the
domain of string perturbation theory at finite genus, they may be regarded
as surfaces beyond the domain of perturbation theory at infinite genus.
The source of the divergences at large orders can be identified as a class
of surfaces which may be regarded as a non-perturbative effect in the theory.
Destabilization of the vacuum by non-perturbative effects has been confirmed,
therefore, in an investigation of the regularized path integral, for which the
tachyonic divergences arising at each order [27][28] have been eliminated.

The techniques developed in this investigation may be adapted to superstring
perturbation theory.  Since no regularization of the amplitudes is required,
a larger class of surfaces, and in particular, those of type $O_G$, can be
included in the superstring path integral.
To count effectively closed surfaces in the path integral,  limits involving a
more rapid decrease of $\vert K_n\vert$ than that given in $\S 2$ such as
$${{\epsilon_0}\over {n^q}}~<~\vert K_n\vert~<~{{\epsilon_0^\prime}\over {n^q}}
{}~~~~~~~q \in [1, \infty)
\eqno(43)$$
must also be used.  However, it has been noted in the proof of perturbative
finiteness of the amplitudes at each order [23] that the integral over the
multipliers containing the term
$$\int~{{d^2 K_n}\over {\vert K_n\vert^2~(ln \vert K_n\vert)^5}}
\eqno(44)$$
is finite whenever $\vert K_n\vert \to 0$.
  Thus, for each value of q, the integral over the multipliers
is finite, and it only remains to be determined whether the sum of all of the
contributions from the different surfaces is finite.  Similar considerations
apply to the integrals over the fixed-point variables.

The large-order divergences of bosonic string theory, which have been derived
here from the genus-dependence of the multipliers and fixed-point distances,
therefore, might be eliminated simultaneously with the infrared divergences
through the introduction of supersymmetry.  This demonstrates the advantage of
using the Schottky group formalism, because it provides a deeper insight into
the connection between supersymmetry and large-order perturbation theory.

\noindent{\bf 6. Conclusion}

Divergences in the scattering amplitudes of the closed bosonic string either
arise from the coincidence of vertex operators, the boundary of moduli space
or the sum over the genus in the perturbative expansion of the S-matrix.  The
resolution of the problem of the integration region for the moduli space
integral in the Schottky group parametrization has led to rigorous bounds for
the regularized vacuum amplitude.  The growth of these bounds with respect
to the genus, together with the positivity of the terms at each order,
indicates
a further instablity of the vacuum.  The application of these methods to
superstring theory would follow from an explicit superstring measure.  As the
regularization of the moduli space integral through a cut-off on the length
of closed geodesics on the surface is no longer necessary, a larger class
of effectively closed surfaces will be included in the evaluation of the
superstring path integral.

\vskip  .5in
\centerline{\it Acknowledgements.}

This investigation was initiated in Brandeis University, Waltham.  I wish
to thank Profs. Abdus Salam, Howard Schnitzer and Alberto Verjovsky for
their interest and encouragement.  Writing of the present version of the
manuscript has been completed in the Department of Applied Mathematics and
Theoretical Physics, Cambridge, and I would like to thank Prof. S. W.
Hawking and Dr. G. W. Gibbons for their hospitality.
Financial support was provided by the Atomic Energy Commission, Vienna.
\vfill
\eject
\centerline{\bf REFERENCES}

\item{[1]}  D. Gross and V. Periwal, Phys. Rev. Lett. ${\underline {60}}$
(1988) 2105 -2108
\item{[2]}  S. Mandelstam, `The Interacting String Picture and Functional
Integration',
Pro-
\vskip 1pt
ceedings of the Workshop on Unified String Theories, Institute for
Theoretical
\vskip 1pt
Physics, Santa Barbara, ed. by M. Green and D. Gross
(Singapore: World Scientific,
\vskip 1pt
1986) pp.46 - 102
\vskip 5pt
\item{[3]}  L. Bers, Bull. Lond. Math. Soc. ${\underline 4}$(1972) 257 - 300
\vskip 5pt
\item{[4]}  J. L. Petersen and J. R. Sidenius, Nucl. Phys. ${\underline
{B301}}$ (1988)
247 -266
\vskip 1pt
K. O. Roland, Nucl. Phys. ${\underline {B313}}$ (1989) 432 - 446
\vskip 5pt
\item{[5]}  J. L. Petersen, K. O. Roland and J. R. Sidenius, Phys. Lett.
${\underline {B205}}$ (1988) 262 - 266
\vskip 5pt
\item{[6]}  E. D'Hoker and S. Giddings, Nucl. Phys. ${\underline {B291}}$
(1987) 90 - 112
\vskip 5pt
\item{[7]}  G. Cristofano, M. Fabbrichesi and K. Roland, Phys. Lett.
${\underline {B236}}$
(1990) 159 - 164
\vskip 1pt
 G. Cristofano, M. Fabbrichesi and K. Roland, Phys. Lett. ${\underline{B244}}$
(1990) 397 - 402
\vskip 1pt
G. Cristofano, M. Fabbrichesi and K. Roland, Phys. Lett. ${\underline {B246}}$
(1990) 45-53
\vskip 5pt
\item{[8]}  M. Fabbrichesi and R. Iengo, Phys. Lett. ${\underline {B264}}$
(1991) 319 - 323
\vskip 5pt
\item{[9]}  G. Moore, Phys. Lett. ${\underline {B176}}$ (1986) 369 - 379
\vskip 5pt
\item{[10]}  J. Birman, `The Algebraic Structure of Surface Mapping Class
Groups',
${\underline{Discrete}}$
\vskip 1pt
${\underline {Groups~and~Automorphic~Functions}}$, ed. by J. Harvey
\vskip 1pt
(London: Academic Press, 1977), pp. 163 - 198
\vskip 5pt
\item{[11]}  S. Nag, ${\underline
{The~Complex~Analytic~Theory~of~Teichmuller~Spaces}}$(John Wiley: New
\vskip 1pt
York, 1988)
\vskip 10pt
\item{[12]}  L. Alvarez-Gaume, G. Moore and C. Vafa, Comm. Math. Phys.
${\underline {106}}$
(1986) 1- 40
\vskip 5pt
\item{[13]}  A. A. Belavin and V. Knizhnik, Phys. Lett. ${\underline {B168}}$
(1986) 201-206
\vskip 5pt
\item{[14]}  J. Lehner, ${\underline {Discontinuous~Groups~and~Automorphic
{}~Functions}}$
Mathematical Sur-
\vskip 1pt
veys, Vol. 8 (Providence: American Mathematical Society, 1964)
\vskip 5pt
\item{[15]}  E. D'Hoker and D. H. Phong , Nucl. Phys. ${\underline {B269}}$
(1986) 205 - 234
\vskip 5pt
\item{[16]}  S. Davis, `Configurations of Handles and Classification of
Divergences
in the String
\vskip 1pt
Partition Function' (to be published in Class. Quantum Grav., 1994)
\vskip 5pt
\item{[17]}  C. L. Siegel, ${\underline{Topics~in~Complex~Function~Theory}}$,
Vol. 3
(Wiley: New York, 1973)
\vskip 5pt
\item{[18]}  H. Minkowski, ${\underline{Geometrie~der~Zahlen}}$ (New York:
Chelsea, 1953)
\item{[19]}  R. Bhatia, ${\underline{Perturbation~Bounds~for~Matrix~
Eigenvalues}}$ Research Notes in Mathematics, Vol. 162 (Essex:Longman, 1987)
\vskip 5pt
\item{[20]}  B. Bollobas, J. Lond. Math. Soc. ${\underline {26}}$ (1982) 201 -
206
\vskip  5pt
\item{[21]}  C. Vafa, Phys. Lett. ${\underline{B206}}$ (1988) 421 - 426
\vskip 5pt
\item{[22]}  J. Zinn-Justin, Phys. Rep. ${\underline{70C}}$(1981) 109 - 167
\vskip 5pt
\item{[23]}  N. Berkovits, Nucl. Phys. ${\underline{B408}}$ (1993) 43 - 61
\vskip 5pt
\item{[24]}  S. Davis, Class. Quantum Grav. ${\underline 6}$ (1989) 1791 - 1803
\vskip 5pt
\item{[25]}  L. Sario and  M. Nakai, ${\underline {Classification~Theory~of~
Riemann~Surfaces}}$
\vskip 1pt
(Berlin: Springer-Verlag, 1970)
\vskip 5pt
\item{[26]}  S. Davis, Class. Quant. Grav. ${\underline 7}$ (1990) 1887 - 1893
\vskip 5pt
\item{[27]}  P. Nelson, Phys. Rep. ${\underline{149C}}$ (1987) 337 - 375
\vskip 5pt
\item{[28]}  A. A. Tseytlin, Phys. Lett. ${\underline{B208}}$ (1988) 221 - 227
\vskip 1pt
A. A. Tseytlin, Int. J. Mod. Phys. A. ${\underline 4}$ (1989) 1257 - 1318

\end